\documentclass[prd,10pt,showpacs,twocolumn,nofootinbib,preprintnumbers,aps,superscriptaddress]{revtex4-1} %

\usepackage{amsmath,amssymb,amsfonts,amsthm}
\usepackage{hyperref}
\usepackage{bm,bbm}
\usepackage{graphics}
\usepackage{mathrsfs}

\newcommand{\be}{\begin{equation}}
\newcommand{\ee}{\end{equation}}
\newcommand{\ba}{\begin{eqnarray}}
\newcommand{\ea}{\end{eqnarray}}
\def\bs{\begin{subequations}}
\def\es{\end{subequations}}
\def\a{\alpha}
\def\b{\beta}

\def\la{\lambda}
\def\k{\kappa}
\def\e{\epsilon}

\def\Om{\Omega}
\def\om{\omega}

\def\vr{\varrho}

\def\cL{{\cal L}}

\def\cM{{\cal M}}

\def\cK{{\cal K}}
\def\cV{{\cal V}}

\def\p{\partial}

\def\x{ q}
\def\ds{d_{\rm S}}
\def\dh{d_{\rm H}}
\newcommand{\Eq}[1]{(\ref{#1})}
\def\lp{\ell_{\rm Pl}}
\def\rme{e}
\def\rmd{d}
\def\rmi{i}
\def\bp{\bar{\partial}}
\newcommand{\oarX}[1]{\href{http://arxiv.org/abs/#1}{{\ttfamily #1}}}
\newcommand{\arX}[1]{\href{http://arxiv.org/abs/#1}{{\ttfamily arXiv:#1}}}
\newcommand{\doin}[2]{\href{http://dx.doi.org/#1}{#2}}

\begin{document}

\title{Fractional and noncommutative spacetimes}

\author{Michele Arzano}
\email{m.arzano@uu.nl}
\affiliation{Institute for Theoretical Physics and Spinoza Institute, Utrecht University, Leuvenlaan 4,
Utrecht 3584 TD, The Netherlands}
\author{Gianluca Calcagni}
\email{calcagni@aei.mpg.de}
\affiliation{Max Planck Institute for Gravitational Physics (Albert Einstein Institute)
Am M\"uhlenberg 1, D-14476 Golm, Germany}
\author{Daniele Oriti}
\email{doriti@aei.mpg.de}
\affiliation{Max Planck Institute for Gravitational Physics (Albert Einstein Institute)
Am M\"uhlenberg 1, D-14476 Golm, Germany}
\author{Marco Scalisi}
\email{marco.scalisi@aei.mpg.de}
\affiliation{Max Planck Institute for Gravitational Physics (Albert Einstein Institute)
Am M\"uhlenberg 1, D-14476 Golm, Germany}
\affiliation{Dipartimento di Fisica e Astronomia, Universit\`a di Catania and}
\affiliation{INFN\ Sezione di Catania, Via Santa Sofia 64, I-95023 Catania, Italy}

\date{July 26, 2011}

\begin{abstract}
We establish a mapping between fractional and noncommutative spacetimes in configuration space. Depending on the scale at which the relation is considered, there arise two possibilities. For a fractional spacetime with log-oscillatory measure, the effective measure near the fundamental scale determining the log-period coincides with the nonrotation-invariant but cyclicity-preserving measure of $\kappa$-Minkowski spacetime. At scales larger than the log-period, the fractional measure is averaged and becomes a power law with real exponent. This can be also regarded as the cyclicity-inducing measure in a noncommutative spacetime defined by a certain nonlinear algebra of the coordinates, which interpolates between $\kappa$-Minkowski and canonical spacetime. These results are based upon a braiding formula valid for any nonlinear algebra which can be mapped onto the Heisenberg algebra. 
\end{abstract}


\preprint{AEI-2011-089}
\preprint{\doin{10.1103/PhysRevD.84.125002}{Phys.\ Rev.\ D {\bf 84}, 125002 (2011)} \qquad [arXiv:1107.5308]}
\maketitle


\section{Introduction}

Despite the advanced level of mathematical tools developed to construct quantum theories of spacetime and geometry, field theory-based models remain the most suitable frameworks wherein to extract effective physics and make predictions that could be of direct relevance for phenomenology. Indeed, approaches such as group field theory \cite{Ori06,Ori11}, loop quantum gravity \cite{Rov06,Thi07}, spin foams \cite{Ori01,Per03}, and simplicial quantum gravity \cite{Wil2,Wil3,lol08,lol2,lol3}, among others, have been accumulating results but they still struggle to get in touch with observations. Part of the difficulty is structural. Several such theories aim at explaining the very origin of the continuum spacetime we are accustomed to, and thus are based on pregeometric, prespacetime, and usually discrete structures from which the usual description of physics has to emerge in terms of continuum spacetime and geometry (and its general relativistic dynamics). Beside the conceptual aspects, this ambitious goal is obviously technically challenging and no setup can claim definite success. This, of course, makes it very difficult also to extract effective dynamical models for both spacetime geometry and matter based on a continuum formalism, and possibly on a field theory language, or to make direct contact with existing effective frameworks.

Conversely, approaches like asymptotically safe gravity \cite{Nie06,NiR,ReS4,CPR}, Ho\v{r}ava--Lifshitz gravity \cite{Muk10,Sot10}, and noncommutative spacetimes \cite{Madore,Szabo,Connes,ADKLW} (whether seen as fundamental or effective) are framed in a field-theoretical language which is more familiar and under control, and thus closer to phenomenological applications.

In this second group of proposals, noncommutative field theories (NCFT) have been extensively studied for more than a decade.  These models turned out to emerge, in one form or another, in a variety of contexts, from string theory \cite{SW, MST}  and loop quantum gravity \cite{ACSS, FKGS} to heuristic approaches to ``quantum spacetime'' \cite{Doplicher:1994zv, Doplicher:1994tu} and ``deformations'' of relativistic symmetries \cite{Majid:1994cy, Agostini:2003vg}. Perhaps, the most striking appearance of a noncommutative space is in the less exotic context of {\it classical} $(2+1)$-dimensional Einstein gravity coupled to point particles \cite{Matschull:1997du}. Since the theory is topological, the latter are introduced as topological defects on the spacetime manifold \cite{Deser:1983tn}. Gravitational interactions turn out to affect profoundly the structure of the particle phase space, leading to group-valued momenta and a nonzero Poisson bracket between the particle spacetime coordinates. When one considers the corresponding field theory, plane waves labelled by group-valued momenta naturally lead to a noncommutative field theory in configuration space \cite{Matschull:1997du, ArzanoAlesci}. Because of this feature, this class of noncommutative geometries is also the most likely one to be related to approaches like loop quantum gravity, spin foam models, or group field theory, heavily involving group-theoretic structures describing quantum spacetime at the fundamental level.

Another field theory aiming to unify different approaches under the same phenomenology, but which is much more recent than NCFTs, can be formulated in fractional spacetimes \cite{fra1,fra2,fra3,fra4,frc1,frc2}. These are continuum models where the measure in the action is not the usual $D$-dimensional Lebesgue measure $\rmd^Dx$, but a Lebesgue--Stieltjes measure $\rmd\vr(x)$ whose form is dictated by arguments taken from fractal geometry. If spacetime is assumed not to be a continuous differentiable manifold but a fractal, then the measure distribution must be modified accordingly. If, moreover, one desires to keep a differentiable structure (which is absent in the most general disconnected fractal constructions), the measure and the differential structure can be determined by the rules of fractional calculus under certain regimes. These models are interesting also because they could be emerging from the pregeometric, quantum structures identified by quantum gravity approaches, as an intermediate stage before the usual smooth spacetime is recovered.

Exotic measures in effective actions can arise in quite different contexts, but a striking similarity was noticed between fractional measures in a certain limit \cite{fra4,frc2} and the configuration-space, cyclic-invariant measure of $\k$-Minkowski spacetime \cite{AAAD}. Also, the spectral dimension of $\k$-Minkowski spacetime turns out to be smaller than 4 in the ultraviolet \cite{Ben08}, giving a further hint that noncommutative spacetimes may show ``fractal'' features. However, the relation between these features and noncommutativity of the coordinates is presently unclear, and information on the spectral dimension alone is not sufficient to describe the geometry of a non-Riemannian space \cite{frc1,frc2}. Also, it would be interesting to know whether the correspondence between the $\k$-Minkowski measure and a particular limit of fractional measures extends to the large class of real-order fractional measures, which are power-law distributions in the coordinates. 

The aim of this paper is to discuss for the first time the relation between spacetimes based on fractional calculus and on noncommutative differential calculus. As anticipated, we focus here only on some limited aspects of the definition of an effective field theory, that is the form of the effective measure. Thus, we focus on (some of) the geometric properties of such spacetimes only. We do not touch on their symmetry structures nor on the kinematical space of fields or on their actual dynamics. The correspondence sketched in \cite{fra4,frc2}, however, will be studied at length and we shall provide a general mapping between real-order fractional measures and noncommutative spaces. 

The resulting physical picture is the following. 
\begin{enumerate}
\item {\it Noncommutative spaces with fractal properties.} 
One of the tools of noncommutative field theories consists in extracting an effective measure for an action defined on the \emph{classical}, commutative configuration space. Certain noncommutative (and, in general, nonlinear) algebras give rise to effective measures which correspond to certain classical fractal spacetimes. Thanks to the correspondence between the effective measure in the noncommutative actions and in this class of fractal theories, we can study in detail these noncommutative theories in the language of fractal geometry.
\item {\it Fractal spaces.} In fractal field theories, classical spacetime is conjectured to have a genuine fractal  structure, perhaps totally disconnected. At some fundamental length scale $\ell_\infty$, which we identify with the Planck length $\lp$, one can approximate this structure via a complex fractional measure, and a formalism on the continuum becomes available. Because of the presence of discrete symmetries at ultramicroscopic scales, the geometry of these spaces is radically different from that of ordinary smooth spaces, up to the point where one may question the distinction between classical and quantum. This is in agreement with the existence of the mapping with noncommutative theories: quantum and fractal properties are intimately related, and one can describe the same space in different languages.
\item {\it Scale hierarchy.} Near $\ell_\infty= \lp$, the geometry of spacetime is $\kappa$-Minkowski. At larger scales, the coordinates obey a nonlinear ``fractional'' noncommutative algebra of order $\a$, corresponding to a fractal geometry with anomalous measure and noninteger Hausdorff and spectral dimensions. Different values of the real parameter $\a$ constitute a multifractal regime which interpolates between $\lp$ and macroscopic scales, where ordinary commutative (or canonical, depending on the details of the fractional algebra) four-dimensional spacetime is recovered. 
\end{enumerate}

We develop this scheme first by reviewing some basic results in the recently proposed fractional spacetimes (Sec.~\ref{frsp}) and in noncommutative theories (Sec.~\ref{ncsp}). Then, we obtain a braiding formula essential for dealing with generic nonlinear noncommutative algebras of coordinates (Sec.~\ref{braid}). This formula is crucial for our main twofold result, presented in Sec.~\ref{nfs}: (i) the mapping between fractional spaces at a certain scale and $\k$-Minkowski spacetime and (ii) the explicit relation between real-order fractional spaces and a new specific type of nonlinear noncommutative algebra. Section \ref{conc} collects some concluding remarks.


\section{Multifractional spacetimes}\label{frsp}

When dealing with fractal sets \cite{Fal03}, ordinary differentiability is given up in favour of other geometric tools. In fact, a fractal (or a multifractal, if its dimension changes with the scale \cite{Har01}) is a set too ``irregular'' to be described by smooth geometry, and even the notion of continuum must be replaced by a description in terms of discrete symmetries. Under certain conditions, however, random and deterministic fractals admit a continuum approximation based upon fractional calculus \cite{Pod99,KST}; this approximation scheme \cite{RLWQ,NLM} was reviewed in \cite{frc1}. Here it is sufficient to recall that, for all practical purposes in physical applications, sets described by fractional operators can be also regarded as full-fledged multifractals (rather than approximations of certain fractals), because they share the main properties by which fractals are characterized. Namely, they are endowed with a discrete scale symmetry \cite{fra4,frc2} and their dimension(s) can be noninteger \cite{fra4,frc1} and varying with the scale \cite{fra4,frc2}.


\subsection{Definition}

Multifractional Euclidean \cite{frc1} and Minkowski \cite{frc2} spaces are the simplest applications of multifractal geometry to spacetime itself. In the following we refer to Refs.\ \cite{fra4,frc1,frc2} for all details. The main building block is fractional Minkowski spacetime $\cM_\a^D$ of order $\a$, defined by an embedding Minkowski spacetime $M^D$ with $D$ topological dimensions, some calculus rules ${\rm Calc}^\a=\{\p^\a, I^\a,\rmd^\a \}$ (for derivatives $\p^\a$, integrals $I^\a$, and external differentials $\rmd^\a$), a complex measure $\tilde\vr_\a$ with a given support, a natural norm $\| \cdot \|$, and a Laplacian $\cK$:
\be
\cM_\a^D = (M^D,\,{\rm Calc}^\a,\,\tilde\vr_\a,\,\| \cdot \|,\,\cK)\,.
\ee
We first specify the complex measure. A complex measure $\vr$ is such that $\vr(\emptyset)=0$, $\vr(\cup_n U_n)=\sum_n\vr(U_n)$ for a sequence of disjoint sets, and $\vr(U)\in\mathbb{C}$. On the other hand, a real measure (or simply a measure) takes only real nonnegative values, $\vr(U)\geq 0$, while the output of a signed measure is a real number but of any sign. $\tilde\vr_\a$ is characterized by a real parameter $0\leq\a\leq1$ and a set of nonnegative modes $\om\geq 0$, such that $\tilde\vr_\a$ is a linear combination of complex measures $\vr_{\a,\om}$, $\tilde\vr_\a=\sum_\om\vr_{\a,\om}$. For each mode $\om$, the measure can be specified by complex fractional calculus. Without loss of generality, one can take the support of the measure to be the positive semiaxis for each direction $x=x^\mu$, so that $\vr_{\a,\om}$ is a linear combination of power laws with complex exponents:
\be
\vr_{\a,\om}= \frac{x^\a}{\Gamma(\a+1)}+A \frac{x^{\a+\rmi\om}}{\Gamma(\a+\rmi\om+1)}+A^* \frac{x^{\a-\rmi\om}}{\Gamma(\a-\rmi\om+1)}\,.
\ee
The complex coefficient $A$ can be chosen to be real, so that the measure is real-valued (then, we say the measure is self-conjugate). Writing $x^{\rmi\om}=\exp(\rmi\om\ln x)$ and noting that
\ba
\frac{1}{\Gamma(\a\pm\rmi\om)} &=&{\rm Re}\left[\frac{1}{\Gamma(\a+\rmi\om)}\right]\pm\rmi {\rm Im}\left[\frac{1}{\Gamma(\a+\rmi\om)}\right]\nonumber\\
&=:& R_\Gamma(\a+\rmi\om)\pm\rmi I_\Gamma(\a+\rmi\om)\,,
\ea
for each direction in spacetime we eventually have
\ba
\vr_{\a,\om}(x) &=& \frac{x^\a}{\Gamma(\a+1)}\left[1+A_{\a,\om}\cos\left(\om\ln\frac{x}{\ell_\infty}\right)\right.\nonumber\\
&&\left.+B_{\a,\om}\sin\left(\om\ln\frac{x}{\ell_\infty}\right)\right]\,,\label{kcom2}
\ea
where
\ba
A_{\a,\om} &=& 2A\, \Gamma(\a+1) R_\Gamma(\a+\rmi\om+1)\,,\\
B_{\a,\om} &=& 2A\, \Gamma(\a+1) I_\Gamma(\a+\rmi\om+1)
\ea
are real and $\ell_\infty$ is a fundamental scale introduced to make the arguments of the logarithmic functions dimensionless. 

At the classical level, the fact that the measure has support only in the orthant $x^\mu\geq 0$ does not lead to any consequence, at least at large-enough scales and away from the boundary (in the bulk, microphysics is described by local equations). The extension to the whole Minkowski embedding can be done only in the infrared limit of integer calculus. At the quantum level, the boundary affects the vacuum state(s) of the theory like, e.g., in the Unruh effect.  These aspects have not been studied in \cite{frc1,frc2} and should deserve further attention.

Notice that $\vr_{\a,\om}$ is a signed measure but not a measure, because of the oscillations. Adding points to a set may result in a decrease of the output value of a signed measure. Then, neither volumes nor the Hausdorff dimension can be defined. In this sense, the oscillatory regime is ``pregeometric,'' even if a geometry does exist. This is not an issue since, on one hand, complex and signed measures are well-defined mathematical objects and, on the other hand, the notions of volume and Hausdorff dimension admit suitable extensions to oscillating measures \cite{frc2}. In the following, we shall continue to call $\tilde\vr_\a$ (and its multifractional generalization) a ``measure'' with this distinction in mind. Denote with $\a_0$ the fractional parameter (or ``fractional charge'') attached to the time direction, and assume that spatial coordinates have the same charge $\a$. In general, $\a_0\neq\a$; if $\a_0=1$, there are no oscillations in time. The logarithmic oscillations are governed by a dimensionless scale 
\be\label{scale0}
\la_\om=\exp(2\pi/\om)\,.
\ee
The oscillatory part of Eq.~\Eq{kcom2} is log-periodic under the discrete scaling transformation 
\be
\ln\frac{x}{\ell_\infty}\,\to\, \ln\frac{x}{\ell_\infty}+\frac{2\pi n}{\om}\,,\qquad n=0,\pm1,\pm2,\dots\,,
\ee
implying an invariance (up to an overall rescaling of $\vr_{\a,\om}$) under the dilation
\be\label{dsi}
x\,\to\, \la_\om^n x\,,\qquad n=0,\pm1,\pm2,\dots\,.
\ee
The characteristic (as opposed to fundamental) physical scale associated with $\la_\om$ is
\be
\ell_\om=\la_\om\ell_\infty>\ell_\infty\,.
\ee

The next step is to sum over all $\om$, which gives the measure on $\cM_\a^D$. However, we are mainly interested in multifractional Minkowski spacetime $\cM_*^D$, where one also sums over all possible $\a$'s. At any given scale, a ``snapshot'' of $\cM_*^D$ shows the structure of $\cM_\a^D$, and as the scale changes so does $\a$. The range of $\a$ can be determined by requiring that the natural norm on the ``snapshots'' $\cM_\a^D$ always respects the triangle inequality, which implies $1/2\leq\a\leq 1$. The total integration measure reads
\be\label{osr}
\rmd\vr(x) = \sum_\a g_\a\sum_\om\prod_\mu \rmd\vr_{\a,\om}(x^\mu)\,,
\ee
where $g_\a$ are some dimensionful coefficients and the sum over $\a$ may also mean integration. Equation \Eq{osr} is the definition of multifractional measure. The differential structure of (multi)fractional spaces is determined by the choice of calculus or, in other words, of the derivative operator. In the presence of a nontrivial metric, one would also have a determinant factor in front of $\prod_\mu \rmd\vr_{\a,\om}(x^\mu)$, with implicit arbitrary dependence on the coordinates. Metric and Lebesgue--Stieltjes measure structures are independent, the former being specified by a two-form $g_{\mu\nu}$ and the latter by a differential structure \cite{frc1}. As one can see in the tetrad formalism, the total measure (i.e., the volume form) is made of both ingredients just like in ordinary geometry: the calculus and the metric determine, respectively, the external differential and the contraction rule of two vielbeins.

The calculus of variations immediately shows that in a given theory there appear two different derivatives, one depending on the boundary at $x^\mu=0$ and the other on the boundary at $x^\mu=+\infty$. We have showed elsewhere \cite{frc1,frc2} that the above measure is associated with the left Caputo derivative
\be
(\p^\a f)(x) := \frac{1}{\Gamma(1-\a)}\int_{0}^{x} \frac{\rmd x'}{(x-x')^\a}\p_{x'}f(x')\,,\label{rcd}
\ee
and the Weyl derivative
\be
({}_\infty\bp^\a f)(x) :=-\frac{1}{\Gamma(1-\a)}\int_{x}^{+\infty}\frac{\rmd x'}{(x'-x)^\a}\p_{x'}f(x')\,.\label{rcd2}
\ee
The functions $f$ belong to the space of absolutely continuous functions on $\mathbb{R}_+$. From these definitions, one can construct external differentials and forms just like in ordinary calculus. We show these expressions for completeness, although we shall not need them in the present work. The Laplacian $\cK$ is a second-order operator in fractional or integer derivatives, depending on the formulation of the theory.

A field theory on multifractional spacetime is described by the action
\be\label{genac}
S=\int \rmd \vr(x)\,\cL\,,
\ee
where all coordinates with fractional charge $\a\neq 1$ run from 0 to $+\infty$, while the time coordinate runs over the whole axis if $\a_0=1$. From the perspective of field theory, the coefficients $g_\a$ in Eq.~\Eq{osr} are coupling constants attached to different operators, and the total multifractional action coincides with what one would get from renormalization group arguments \cite{frc2}.


\subsection{Properties and scale hierarchy}

Multifractional field theories undergo a sequence of regimes according to the scale $\ell$ probed by the observer \cite{fra4,frc2}. At scales $\ell< \ell_\infty$, the nonsmooth geometric and topological structure of the underlying multifractal begins to emerge (what we called ``boundary effects'' in \cite{frc2}) and the continuum picture breaks down.
Starting from scales just at this threshold, $\ell\sim \ell_\infty$, we can employ the fractional formalism. Expanding Eq.~\Eq{kcom2} around $x/\ell_\infty\sim 1$ and summing over $\a$ and $\om$ we have, up to a finite normalization constant,
\be\label{noncor}
\rmd\vr(x)\sim \rmd^D x\,v_{\rm BE}(x):=\rmd^D x\, t^{-\e_0} \prod_{j=1}^{D-1}x_j^{-1}\,,
\ee
where $\e_0=1$ if $\a_0\neq 1$ and $\e_0=0$ if $\a_0=1$. The subscript BE stands for ``boundary effect'' regime.

In the range $\ell_\infty<\ell\ll\ell_*$, where $\ell_*$ is some other scale specified below, one should take the full form of Eq.~\Eq{osr}. The Hausdorff and spectral dimensions are defined as averaged quantities over a logarithmic period \cite{frc2}. The symmetry of the theory is Eq.~\Eq{dsi}, called discrete scale invariance (DSI), which appears in a number of chaotic systems with fractal properties \cite{Sor98}. Despite the model being continuous, the presence of a DSI at small scales makes it an interesting candidate for an effective description of a discrete-to-continuum spacetime transition. To get the continuum limit in a formal way, one should send the frequency to zero from above, so that the length cutoff vanishes: $\ell_\om\to 0$ as $\om\to 0^+$. However, at mesoscopic spacetime scales much larger than the period, $\ell_\om\ll\ell\lesssim \ell_*$, one can take the average of the measure \cite{frc2},
\be\label{mfrm}
\vr_\a(x):= \langle\vr_{\a,\om}(x)\rangle\propto \prod_\mu (x^\mu)^\a\,,
\ee
so that the effective measure weight in $\rmd\vr_\a(x)=\rmd^Dx\,v_\a(x)$ is
\be\label{avev}
v_\a(x) = t^{\a_0-1}\prod_{i=1}^{D-1} x_i^{\a-1}\,,
\ee
up to a proportionality constant. The average of the oscillations is zero and one remains only with the zero mode. This corresponds to randomizing the fractal structure. The total integration measure is
\be
\rmd\vr(x)\sim \sum_\a g_\a\,\rmd\vr_\a(x)\,,\qquad \ell_\om\ll\ell\lesssim \ell_*\,.\label{mufrr}
\ee

The Euclidean volume of a $D$-ball of radius $R$ scales as $\cV\sim R^{\dh}$, where $\dh$ is the Hausdorff dimension of spacetime. For a fixed $\a$, 
\be\label{dh}
\dh=\a_0+(D-1)\a\,.
\ee
On the other hand, the spectral dimension (i.e., the dimensionality felt by a test particle diffusing in fractional spacetime) is
\be\label{ds}
\ds=D\a\,,
\ee
which is smaller than or equal to $\dh$ if $\a\leq\a_0$. If $\a>\a_0$, fractional spacetime cannot be considered a fractal \cite{frc1}.

In this phase, continuous symmetries emerge. The zero mode in the measure (the surviving part after averaging) breaks ordinary Poincar\'e invariance, but it is invariant under nonlinear transformations of the embedding coordinates $x^\mu$ which preserve the fractional line element \cite{frc1,frc2}. Defining the geometric coordinates $\x^\mu(x):=\vr_\a(x^\mu)$, the measure $\vr_\a(x)$ is invariant under
\be\label{fpotra}
{\x'}^\mu(x) = \tilde\Lambda_\nu^\mu \x^\nu(x)\,,\qquad \x^\mu(x):=\frac{(x^\mu)^\a}{\Gamma(\a+1)}\,,
\ee
where $\tilde\Lambda_\nu^\mu$, satisfying $\tilde\Lambda_\nu^\mu\tilde\Lambda_\mu^\rho=\delta_\nu^\rho$, are Lorentz matrices. If $\a_0\neq\a$, there is a length scale hidden in these matrices. These symmetries, or their ordinary Lorentz version with $\a=\a_0=1$, can be imposed to the Lagrangian density $\cL$ in order to define a field theory and constrain the allowed operators.

Clearly, this modification of standard flat spacetime symmetries is another important aspect of fractional spaces, with potential phenomenological consequences. Deformations of Poincar\'e symmetries also characterize noncommutative spacetimes and are the basis of much of their associated phenomenology. An important direction of work in trying to relate these two effective quantum gravity frameworks would be to do so at the algebraic level of their characteristic symmetries. We do not pursue this direction here, but we stress it would be complementary to our approach and similar in motivations. 

In a multifractional setting, the dimension and the symmetries change with the scale, implicit in $\a=\a(\ell)$. 
The UV and infrared Hausdorff dimensions of spacetime are tightly related to each other. When $\a_0=\a$, the theory has a critical point at $\a=\a(\ell_*)=2/D$, corresponding to $\dh=2$. If $\a(\ell_*)$ is also the lowest possible $\a$, where dimensional flow stops, then one must have $D=4$, otherwise the triangle inequality would be violated during the flow. Thus, four dimensions are selected by geometry arguments. This is no longer true if $\a_0=1$. In general, for a given topological dimension $D\geq 1$, not all fractional measures are possible. Hausdorff dimensions $\dh=\a_0+(D-1)\a$ are associated with well-defined norms only if $\a,\a_0\geq 1/2$. If $\a_0=\a$, this implies \cite{frc2}
\be
D\leq 2 \dh\,,
\ee
while for $\a_0=1$ one has
\be\label{dstar}
D\leq 2 \dh-1\,.
\ee

Finally, at scales much larger than the characteristic scale $\ell_*$ at which the UV critical point is attained, ordinary field theory on Minkowski spacetime is recovered, $\vr(x)\sim \vr_{1}(x)=x$. The theory is Poincar\'e invariant in the standard sense, and the Hausdorff and spectral dimensions of spacetime are close to the topological dimension, $\dh=\ds=D-\e$, where $D=4$ and $\e\ll 1$.


\section{Noncommutative spacetimes}\label{ncsp}

After this brief overview of multifractional geometries, we introduce the basics of noncommutative spacetimes. These spacetimes emerge in several contexts as a practical tool to go beyond the realm of ordinary (quantum) field theory. In the spirit of noncommutative geometry \cite{ADKLW, Connes, Madore}, the algebra of functions on spacetime, including {\it coordinate functions}, becomes the central object.  As a result, the picture of the underlying spacetime arena as an ordinary differentiable manifold is now lost, and it becomes meaningful only in the limit in which the ordinary Abelian product on the algebra of functions is recovered.

The introduction of the non-Abelian $*$-product is reminiscent of the Moyal product in the Weyl quantization approach \cite{Weyl,Gutt} to ordinary nonrelativistic quantum mechanics.  Indeed, the study of field theories on noncommutative spacetimes is greatly simplified by the introduction of a {\it Weyl map} \cite{MSSW, Kosinski:1999dw,Agostini:2002de}, which is nothing but the generalization to field theory of the linear map from the classical phase-space functions to functions of quantum operators, first used in Weyl quantization. In other words, this map encodes the choice of quantization prescription for noncommuting variables, and thus it determines an operator ordering. As such, it is not invertible strictly speaking (there are many quantum theories for any given classical theory). However, by introducing the noncommutative $*$-product on a space of functions of classical coordinates, one defines a new space of such functions that can be put in one-to-one correspondence with the space of functions of quantum operators, with the $*$-product reproducing, in the space of functions of commuting variables, the quantum noncommutativity of functions of operators. If one then interprets the Weyl map as acting on such extended space of classical functions endowed with a noncommutative $*$-product, the map becomes invertible. Let $\Om$ be an invertible Weyl map. Focusing on plane waves (the basic building blocks of a field theory), $\Om$ relates classical plane waves\footnote{They are classical in the sense of being functions of commuting coordinates, even though they themselves multiply via star multiplication.} $\rme^{i k\cdot x}$ and noncommutative plane waves, generically denoted as $w_k(X)$:
\be\label{Weyl}
\Om(\rme^{i k\cdot x})=w_k(X)\,,
\ee
where $k\cdot x=k_\mu x^\mu=-k^0 t+\cdots+k_{D-1} x^{D-1}$ in $D$ topological dimensions [we use the mostly-plus signature $(-,+,\dots,+)$], $x$ are classical coordinates of the space equipped with the $*$-product, $k$ are momenta, and $X$ are elements of the noncommutative algebra.

The $*$-product is defined on the composition of plane waves, so that
\be\label{sta1}
e^{i k\cdot x} * e^{i p\cdot x}:=\Om^{-1}\left[w_k(X) w_p(X)\right]\,.
\ee
In general, to construct a field theory on noncommutative spacetimes one must define a spectral theory via the eigenfunctions of the generators of translations, that is, the plane waves themselves. In other words, one needs an invertible transformation from configuration to momentum space and to know how plane waves compose. If plane waves are phases, this corresponds to finding the non-Abelian composition law for momenta associated with the exponentiated elements of the $*$-algebra. On the classical and on the noncommutative space one will have, respectively,
\bs\label{sta2}\ba
&& e^{i k\cdot x} * e^{i p\cdot x} = C(k,p)e^{i \gamma(k,p)\cdot x}\,,\\
&& w_k(X) w_p(X)=C(k,p)\, w_{\gamma(k,p)}(X)\,,%
\ea\es
where $\gamma^\mu(k,p)$ is a vector function of the momenta determined by the algebra and $C(k,p)$ is some momentum-dependent function. The problem will be to find a mapping $\Omega$ realizing Eqs.~\Eq{sta1} and \Eq{sta2}.  As we will see in the examples below, in certain cases $\gamma^\mu(k,p)$ and $C(k,p)$ can be calculated once a normal ordering prescription for the factors of noncommuting plane waves is given. In this case, 
the calculation boils down to a derivation of a {\it braiding relation} of the type
\be\label{tx}
\rme^{\rmi k_0 T} \rme^{\rmi p_j X^j} \stackrel{?}{=}\rme^{g(X_i,p_i,k_0)} \rme^{\rmi k_0 T}\,,
\ee
where Latin indices run over spatial directions and $g$ is some function of the spatial coordinate operators, the associated momenta, and $k_0$. Before embarking ourselves in the detailed calculations it will be of help to recall a particularly useful integral representation of the Baker--Campbell--Hausdorff (BCH) formula \cite[theorem 5.5]{Mil72}. We need the special function
\be\label{psi}
\psi(z):=\frac{z\ln z}{z-1}=1-\sum_{n=1}^{+\infty}\frac{(1-z)^n}{(n+1)n}\,,
\ee
which is called the generating function for the Bernoulli numbers $B_n$. In fact,
\be\label{bern}
\psi(\rme^{-y})=\frac{y}{\rme^y-1}=\sum_{n=0}^{+\infty}B_n\frac{y^n}{n!}\,,
\ee
where we use the minus-sign convention for $B_1$: $B_0=1$, $B_1=-1/2$, $B_2=1/6$, $B_3=0$, $B_4=-1/30$, and so on.
Also, given an operator $\tilde T$ in an algebra, the linear adjoint action ${\rm ad}_{\tilde T}$ through $\tilde T$ is ${\rm ad}_{\tilde T} \tilde X:=[\tilde T,\tilde X]$, for any $\tilde X$ in the algebra. Thus, one can use the notation
\be\label{adj}
\rme^{{\rm ad}_{\tilde T}} \tilde X:= \sum_{n=0}^{+\infty}\frac{1}{n!}\underbrace{[\tilde T,[\tilde T,\cdots [\tilde T}_{\textrm{$n$ times}},\tilde X]]\cdots]
\ee
to indicate an infinite sum of nested commutators. By using some elementary properties of composite operators, it is not difficult to prove that the BCH formula can be written as
\be\label{BCH}
\ln\left(\rme^{\tilde T} \rme^{\tilde X}\right)=\tilde T+\int_0^1\rmd s\,\psi\left(\rme^{{\rm ad}_{\tilde T}} \rme^{s{\rm ad}_{\tilde X}}\right)\,\tilde X\,.
\ee
This formula should be intended as a very compact notation for a multiple series of nested commutators (acting on $\tilde X$) given explicitly by the binomial series and the series expansions of Eqs.~\Eq{psi} and \Eq{adj}, everything under integration. For algebras of the form 
\be\label{falg}
[\tilde X,\tilde T]=\rmi\la F(\tilde X)\,,
\ee
where $\la>0$ is a constant, Eq.~\Eq{BCH} drastically simplifies. In fact, the action of $\rme^{s{\rm ad}_{\tilde X}}$ is trivial and one can drop both that operator and the integration. Then, using Eq.~\Eq{bern}, we arrive at
\ba
\ln\left(\rme^{\tilde T} \rme^{\tilde X}\right)&=& \tilde T+\psi\left(\rme^{{\rm ad}_{\tilde T}}\right)\,\tilde X\nonumber\\
&=&\sum_{n=1}^{+\infty}B_n\frac{(-1)^n}{n!}\underbrace{[\tilde T,[\tilde T,\cdots [\tilde T}_{\textrm{$n$ times}},\tilde X]]\cdots]\nonumber\\
&&+\tilde T+\tilde X\,.\label{BCH2}
\ea
To compare this relation with its counterpart where $\tilde T$ and $\tilde X$ are exchanged in the left-hand side, one should recalculate Eq.~\Eq{BCH}; however, now the above simplifications do not take place. The procedure can be quite difficult for a general algebra \Eq{falg}, and Eq.~\Eq{tx} seems a hard goal to achieve. On top of that, it is not at all guaranteed that the most convenient Weyl mapping from classical to operator space will select phases as the natural plane waves. This is true, as we will see below, for canonical and $\kappa$-Minkowski spacetimes, where the direct verification of Eq.~\Eq{tx} is straightforward, but it may not be the case for the nonlinear algebras we shall consider. Therefore, we will adopt a very economic strategy based upon the simple properties of canonical noncommutative spacetimes. To begin with, let us discuss briefly the two most studied examples of noncommutative spacetimes, the canonical or Moyal spacetime and $\kappa$-Minkowski spacetime.


\subsection{Canonical spacetime}

Moyal space is the simplest example of noncommutative spacetime. The commutator of spacetime coordinates can be seen as a generalization of the Heisenberg algebra
\be
[X_{\mu}, X_{\nu}]= i \theta_{\mu\nu}\,, 
\ee  
where $\theta_{\mu\nu}$ is a constant antisymmetric matrix.  Interest in this type of noncommutative spacetime was at first triggered by its appearance in the context of string theory \cite{SW,MST} and some years later for its relation to a quantum deformation of the Poincar\'e algebra in which $\theta_{\mu\nu}$ plays the role of a deformation matrix \cite{Chaichian:2004yh, Balachandran:2007vx}.

We focus here on the case in which only the space/time components of the deformation matrix are nonvanishing and are given by $\theta_{0i}=-\lambda$. For this particular choice, we will call $Q_i$ the spatial coordinates. The spacetime algebra is
\be\label{cano}
[Q_i,T]=\rmi\la\,,\qquad [Q_i,Q_j]=0\,,
\ee
with $X_0=T$, which corresponds to the case $F(Q)=1$ in Eq.~\Eq{falg}. Given a function $f(x)$ in ordinary spacetime and its Fourier transform
\be
\tilde f(k)=\int_{-\infty}^{+\infty}\rmd^D x\,\rme^{-\rmi k\cdot x} f(x)\,,
\ee
the time-to-the-right canonical Weyl map is defined as
\be\label{caom}
\Om_q(f):=\frac{1}{(2\pi)^D}\int_{-\infty}^{+\infty}\rmd^D k\,\rme^{\rmi k_j Q^j}\rme^{-\rmi k_0 T}\tilde f(k)\,.
\ee
To calculate the composition law of plane waves, define 
\be\label{tildes}
\tilde T:=-\rmi k_0 T\,,\qquad \tilde Q_j=\rmi p_j Q_j\,,
\ee
so that the new coordinates obey $[\tilde Q_j,\tilde T]=\rmi\tilde \la_j$, where
\be
\tilde\la_j:=k_0 p_j\la\,.
\ee
For this canonical algebra and for a fixed $j$, application of the BCH formula is straightforward. From Eq.~\Eq{BCH2}, one has
\be\nonumber
\exp(\tilde T)\,\exp(\tilde Q_j)=\exp(\tilde T+\tilde Q_j)\,\exp\left(-\frac{\rmi\tilde\la_j}{2}\right)\,.
\ee
Switching $\tilde T$ with $\tilde Q_j$ in the left-hand side of Eq.~\Eq{BCH} eventually yields the same formula with $\tilde\la_j\to -\tilde\la_j$, so that we obtain
\be\nonumber
\exp(\tilde T)\,\exp(\tilde Q_j) = e^{-\rmi \tilde\la_j}\exp(\tilde Q_j)\,\exp(\tilde T)\,.
\ee
From Eq.~\Eq{tildes} and extending to $D$ dimensions,
\ba
&&\exp(-\rmi k_0 T)\,\exp(\rmi p_j Q^j)\nonumber\\
&&\quad = e^{-\rmi\la k_0\sum_jp_j}\exp(\rmi p_j Q^j)\,\exp(-\rmi k_0 T)\,,\label{braq}
\ea
where upper and lower indices are summed over. Thus, the exchange of time and space exponentials generates an extra constant (with respect to spacetime coordinates) phase. This constant changes for other choices of the operator ordering in Eq.~\Eq{caom} (time-to-the-left and symmetric prescriptions), but one can easily show that the final result \Eq{cycleq1} is unaffected. In this respect, the choice \Eq{caom} does not lead to any loss of generality. Using Eq.~\Eq{braq}, we can combine two noncommutative plane waves and, by the inverse Weyl map, we obtain the following $*$-product:
\ba
e^{i k_\mu q^\mu} *_q e^{i p_\mu q^\mu} &=& \Om^{-1}_q [\exp(ik_jQ^j)\exp(-ik_0T)\nonumber\\
&&\times\exp(ip_jQ^j)\exp(-ip_0T)]\nonumber\\
&=&e^{i (k_\mu+p_\mu) q^\mu}e^{-i\lambda  k_0 \sum_j p_j}\,.\label{starq}
\ea

The action functional of noncommutative field theory is a linear map from noncommutative spacetime to $\mathbb{C}$. In order to remove ambiguities in the definition of interaction terms in the action functional, it is customary to impose the cyclic property \cite{DJMTWW,AAAD}. In the $*$-product formalism, the latter can be written as
\be\label{cycl}
\mathcal{I}([\hat{f},\hat{g}]):=\int\rmd^D x\ v(x)[f,g]_* = 0 \,,
\ee
where $[f,g]_*=f*g-g*f$ and $v(x)$ is a measure weight that assures the requested feature when combined with the star-commutator of any two test functions $f$ and $g$.

In the canonical case, it is well known that the trivial measure $v_c(q)=1$ satisfies Eq.~\Eq{cycl}.
However, simply by imposing condition \Eq{cycl}, one can get a whole class of cyclicity-inducing measures.

Given that the functions $q^n:=q_0^{n_0}\dots q_{D-1}^{n_{D-1}}$ form a basis for any $f$ and $g$, we can rewrite Eq.~\Eq{cycl} as
\bs\label{cycl1}\ba
&&\int\rmd^D q\ v_c(q)[q_j^n,g(q)]_{*_q} = 0\,,\\
&&\int\rmd^D q\ v_c(q)[t^n,g(q)]_{*_q} = 0 \,, \qquad \forall~n~\in \mathbb{N}.
\ea\es
Now, taking $g(q)=e^{ip\cdot q}$, we want to formulate the $*$-product as a pseudodifferential operator starting from the following observation:
\bs\label{stdif}\ba
q_j*_q e^{ip\cdot q}&=&\lim_{k\rightarrow 0}(-i\partial_{k^j}e^{ik\cdot q}*_q e^{ip\cdot q})\nonumber\\
&=&\lim_{k\rightarrow 0}[-i\partial_{k^j}e^{i(k+p)\cdot q}e^{-i\lambda k_0  \sum_j p_j}]\,,\\
t*_q e^{ip\cdot q}&=&\lim_{k\rightarrow 0}(i\partial_{k^0}e^{ik\cdot q}*_q e^{ip\cdot q})\nonumber\\
&=&\lim_{k\rightarrow 0}[i\partial_{k^0}e^{i(k+p)\cdot q}e^{-i\lambda k_0  \sum_j p_j}]\,.
\ea\es
Therefore, we can write the star-commutators as
\bs\label{comsta}\ba
[q_j^n,e^{ip\cdot q}]_{*_q}&=&(-i)^n\lim_{k\rightarrow 0}\p_{k^j}^n[e^{i(k+p)\cdot q}e^{-i\lambda k_0  \sum_j p_j}\nonumber\\
&&-e^{i(k+p)\cdot q}e^{-i\lambda  p_0 \sum_j k_j}]\nonumber\\
&=&\{q_j^n-[q_j-(D-1)\lambda p_0]^n\}e^{ip\cdot q}\,,\\
\ [t^n,e^{ip\cdot q}]_{*_q}&=& i^n\lim_{k\rightarrow 0}\p_{k^0}^n[e^{i(k+p)\cdot q}e^{-i\lambda k_0 \sum_j p_j}\nonumber\\
&&-e^{i(k+p)\cdot q}e^{-i\lambda  p_0 \sum_j k_j}]\nonumber\\
&=&\Big[\Big(t+\lambda \sum_jp_j\Big)^n-t^n\Big]e^{ip\cdot q}\,,
\ea\es
so that
\bs\label{comdif}\ba
[q_j^n,e^{ip\cdot q}]_{*_q} &=&\{(-i\p_{p_j})^n-[-i\p_{p_j}-(D-1)\lambda p_0]^n\}e^{ip\cdot q}\,,\nonumber\\\\
\ [t^n,e^{ip\cdot q}]_{*_q} &=& \Big[\Big(i\p_{p_0}+\lambda \sum_jp_j\Big)^n-(i\p_{p_0})^n\Big]e^{ip\cdot q}\,.\nonumber\\
\ea\es

At this point, we extend the result to any function by linearity with Fourier analysis and, integrating over momenta, we gain the differential form of the star-commutators for any function $g$:
\bs\label{comdifg}\ba
[q_j^n,g(q)]_{*_q}&=&\{q_j^n-[q_j+i(D-1)\lambda \p_t]^n\}g(q)\,, \\
\ [t^n, g(q)]_{*_q}&=&\Big[\Big(t-i\lambda \sum_j\p_j\Big)^n-t^n\Big]g(q)\,.
\ea\es
Substituting these expressions in Eq.~\Eq{cycl1} and integrating by parts, we get the cyclicity-inducing equations
\bs\label{cycleq1}\ba
\p_tv_c(q)=0\,,\\
\label{cycleq2}
\sum_j\p_jv_c(q)=0\,.
\ea\es
Their solutions are all the measures assuring the cyclic property of the canonical action functional. As we could expect, the trivial measure $v_c=1$ is a solution of the latter equations. This particular solution is the measure we have for a commuting spacetime and it coincides with the Lebesgue measure. More general cyclicity-inducing measures $v_c(q)\neq1$ are
\be\label{cmeas}
v_c(\mathbf{q})= \exp\left[ a\sum_{j=1}^{D-1}q_j - \frac{a(D-1)}{b}\ln\left(\sum_{j=1}^{D-1}e^{bq_j}\right)\right]\,,
\ee
for any real $a$ and $b\neq0$.


\subsection{$\kappa$-Minkowski spacetime}

Noncommutative spacetimes in which the coordinates close a Lie algebra are of particular interest. As discussed in the introduction, they emerge in the description of point particles coupled as topological defects to three-dimensional Einstein gravity, and are also the type of effective noncommutative spaces most easily related to or derived from group field theories \cite{eterawinston,noi,danieleEmerg}. In such theories, momentum space is described by the Lie group associated with the spacetime Lie algebra. Moreover, in all known examples there exists a notion of deformed relativistic symmetries in which the parameter providing the dimension of length for the structure constants of the Lie algebra plays the role of a deformation parameter. The particular example of $\kappa$-Minkowski space \cite{Majid:1994cy,LRZ,LNRT} is one of the best studied models of Lie-algebra noncommutative spacetime. It is characterized by the $\kappa$-Poincar\'e algebra \cite{LNRT}, a type of deformation of relativistic symmetries which became popular as an example of relativistic symmetries incorporating an invariant energy scale, and with interesting phenomenological implications \cite{AmelinoCamelia:2000mn, KowalskiGlikman:2002we}.  The commutators for $\kappa$-Minkowski space are given by
\be\label{km}
[X_i,T]=\rmi\la X_i\,,\qquad [X_i,X_j]=0\,,
\ee
where $\la=1/\kappa>0$ has now the dimension of length/inverse energy.  We look again at the composition of plane waves and the braiding relation for noncommuting exponentials. In this case we employ the BCH formula \Eq{BCH2} for
\be\label{tildesk}
\tilde T:=-\rmi k_0 T\,,\qquad \tilde X:=\rmi p_j X^j\,.
\ee
For a $\kappa$-Minkowski algebra, [$F(X)\propto X$ in Eq.~\Eq{falg}], Eq.~\Eq{BCH2} is still fully resummed:
\be\label{use}
\exp(\tilde T)\,\exp(\tilde X)=\exp\left(\tilde T+\frac{\rmi\tilde\la}{\rme^{\rmi\tilde\la}-1}\tilde X\right)\,,
\ee
where
\be
\tilde\la:= -\rmi k_0\la\,.
\ee
Exchanging $\tilde T$ with $\tilde X$ in the left-hand side of \Eq{BCH} and recalculating, one finds Eq.~\Eq{use} with $\tilde T$ and $\tilde X$ flipped in the left-hand side and $\tilde\la\to -\tilde\la$ in the right-hand side; rescaling $\tilde X$, one gets
\be\nonumber
\exp(\tilde T)\,\exp(\tilde X)= \exp\left(\rme^{-\rmi\tilde\la}\tilde X\right)\,\exp(\tilde T)\,,
\ee
so that
\ba
&&\exp(-\rmi k_0 T)\,\exp(\rmi p_j X^j)\nonumber\\
&&\quad =\exp\left(\rmi  \rme^{-\la k_0}p_j X^j\right)\,\exp(-\rmi k_0 T)\,.\label{brak}
\ea
Equipped with this relation, one can derive the explicit group-law addition of momenta from the composition of plane waves. We take a particular choice of normal ordering with the time variables always appearing to the right,
\be\label{R}
\Om_R(f):=\frac{1}{(2\pi)^D}\int_{-\infty}^{+\infty}\rmd^D k\,\rme^{\rmi k_i X^i}\rme^{-\rmi k_0 T}\tilde f(k)\,,
\ee
so that
\ba
e^{i k\cdot x} *_R e^{i p\cdot x}&:=& \Om_R^{-1}[\exp(\rmi k_j X^j)\,\exp(-\rmi k_0 T)\,\exp(\rmi p_j X^j)\nonumber\\
&&\times\exp(-\rmi p_0 T)]\,\stackrel{\Eq{brak}}{=} e^{i \gamma_{R}(k,p)\cdot x}\,,\label{brak2}
\ea
where $\gamma_{R}(k,p)=(k^0+p^0, k^j + \rme^{-\la k^0}p^j)^{\rm t}$ (t denotes transpose). From a deformed-symmetry point of view, such relation implies that the generators of translations (of which plane waves are eigenfunctions) exhibit a nontrivial coproduct \cite{Arzano:2007nx, CamArz}. From a geometrical point of view, Eq.~\Eq{brak2} can be interpreted in terms of the Maurer--Cartan connection naturally defined on the momentum space group manifold \cite{AmelinoCamelia:2011bm, AmelinoArzRelLoc,Arzano:2010jw}.

Fields on $\kappa$-Minkowski space are more difficult to handle than their counterparts defined on canonical spacetimes \cite{CamArz, Arzano:2007ef, Freidel:2007yu, Agostini:2006nc,Arzano:2009ci}.  One difficulty one immediately faces is that the cyclicity property is lost if one assumes the trivial measure $v_\kappa=1$. It is therefore necessary to find the appropriate cyclicity-inducing equations from the general condition \Eq{cycl}. These were found in \cite{AAAD}, following the same procedure we outlined in the previous section. The resulting equations, for the choice of normal ordering just discussed above, are $(1-e^{-in \lambda \p_t})v_\kappa(x)=0$ (where $n$ is an integer) and $\p_j[x^jv_\kappa(x)]=0$, equivalent to
\be\label{cike}
\p_t v_\kappa(x)=0\,,\qquad \p_j[x^jv_\kappa(x)]=0\,.
\ee
Effective measures do not need to respect the same symmetries of the algebra.\footnote{One could also try to obtain symmetry-preserving measures by dropping the cyclicity requirement, at the cost of having to deal with ordering ambiguities in the definition of the Lagrangian density, or try to preserve both by modifying some other ingredient of the theory. This issue will not concern us here, and we will stick to the cyclicity condition and regard the preservation of symmetries to be of secondary importance.} 
In \cite{AAAD}, the rotation-invariant solution $v_\kappa({\bf x})=|\mathbf{x}|^{1-D}$ was chosen, but it is clear we could pick other cyclicity-inducing measures that break rotation symmetry, such as
\be\label{kiacob}
v_\kappa({\bf x})= \prod_{j=1}^{D-1}|x_j|^{-1}\,.
\ee

The time-to-the right normal ordering in Eq.~\Eq{R} is not the only possibility. One can symmetrize the time dependence as
\be\label{T}
\Om_T(f):=\frac{1}{(2\pi)^D}\int_{-\infty}^{+\infty}\rmd^D k\,\rme^{-\frac{\rmi}{2} k_0 T}\rme^{\rmi k_i X^i}\rme^{-\frac{\rmi}{2} k_0 T}\tilde f(k)\,,
\ee
or simply take the symmetric Weyl map
\be\label{S}
\Om_S(f):=\frac{1}{(2\pi)^D}\int_{-\infty}^{+\infty}\rmd^D k\,\rme^{\rmi k_\mu X^\mu}\tilde f(k)\,.
\ee
Explicit lengthy calculations show that these alternatives still yield the same solutions \cite{AAAD}. As we show later, one can find a cyclicity-inducing measure following a different and simpler procedure based on a coordinate transformation.


\section{Canonical mapping\\ and braiding formula}\label{braid}

Heisenberg algebras are well understood and there is a sizable literature about the construction of a field theory on a canonical spacetime \cite{DoNe}. Now we want to show that it is possible to manage a certain class of algebras and their cyclic properties just starting from the canonical one and exploiting what we know about it. This method was first employed in \cite{AAAD} in the case of $\kappa$-Minkowski spacetime, but what follows is a more general formulation valid also for nonlinear algebras.

Let $[X_{\mu}, X_{\nu}]= i \theta_{\mu\nu}(X)$ be an algebra and $S = \int\rmd^D x\ v(x)F(x)$ the associated action integral in the $*$-product formalism.\footnote{
The $*$-product is not used in multiplying a Lagrangian density (itself involving $*$-products of fields and their derivatives) and the measure function. In fact, the measure factor can be absorbed into the definition of a trace functional on the noncommutative algebra, and integration of a function is defined by taking the trace.} Let $Q=(T,\mathbf{Q})^{\rm t}$ obey the relations \Eq{cano} and  $\int\rmd^D q\ v_c(q)f(q)$ be the correspondent action. Suppose there exists an invertible coordinate transformation such that
\be\label{cotr}
X_{\mu}=X_{\mu}(Q)\,.
\ee
Then, we can describe any functional $F(X)$ of the general algebra as a functional $f(Q)=F[X(Q)]$ of the canonical operators. Obviously, we can repeat the same argument for the classical coordinates, $(t,\mathbf{x})$ and $(t,\mathbf{q})$, in the spaces equipped with the $*$-product. Writing down the canonical action functional and operating the coordinate transformation, we have
\be\label{canmap}
\int\rmd^D q\ v_c(q)f(q)=\int\rmd^Dx\ v_c[q(x)] J(x) F(x)
\ee
where 
\be
J(x)=\left|\frac{\p q}{\p x}\right|
\ee
is the Jacobian of the transformation. Equation \Eq{canmap} is just the action functional $S$ if $v(x)=v_c[q(x)]J(x)$.
So, calculating the Jacobian would allow us to find cyclicity-inducing measures of any algebra related to the canonical one. In particular, the Jacobian itself is always a solution because it corresponds to the case $v_c=1$.


\subsection*{Braiding formula}

Equation \Eq{braq} is the simplest example of a braiding formula, a key ingredient in the following calculations. In fact, we can prove the very useful relation
\be\label{brag}
e^{\tilde T} e^{\tilde X(Q)} =e^{\tilde X(Q-\rmi\tilde \la)}e^{\tilde T}\,,
\ee
where $\tilde X(Q)$ is an analytic function of the canonical coordinate $Q$ (later the generalization to $D$ dimensions will be straightforward). To begin, we expand the operator $e^{\tilde X(Q)}$ in its series definition (assuming it exists),
\be\label{eg}
e^{\tilde X(Q)}=\sum_{n=0}^{+\infty} a_n Q^n\,,
\ee
where $a_n$ are some coefficients.\footnote{One might legitimately question whether, for the validity of this argument, such series must be convergent (in particular, $a_n<\infty$ for all $n$) or else if it is sufficient to consider a formal expression. In the examples we are concerned with in this work, however, we can safely ignore this issue since, as we show below, one can slightly modify the operators involved without changing the underlying theory, and make Eq.~\Eq{eg} well defined.} Expanding also $e^{\tilde T}$, the problem is to move $\tilde T$ operators to the right in terms of the form $\tilde T^m Q^n$. Repeatedly using Eq.~\Eq{cano}, one gets
\be
\tilde T^m Q^n = \sum_{l=0}^n l!(\rmi\tilde \la)^l  \binom{n}{l}\binom{m}{l} Q^{n-l} \tilde T^{m-l}\,.
\ee
Terms with $m<l$ vanish (replace binomials with their definition with gamma functions), so that resumming the exponential in $\tilde T$ we obtain
\be\label{bragq0}
e^{\tilde T} Q^n = (Q-\rmi\tilde \la)^n e^{\tilde T}\,,
\ee
and using Eq.~\Eq{eg} we have Eq.~\Eq{brag}. Reinstating momentum factors, extending to $D$ dimensions, and writing $\tilde T=-\rmi k_0 T$ and $\tilde X(Q_j)=\rmi p_j X(Q_j)$ for some $X(Q_j)$ dependent only on the $j$th coordinate, one has
\ba
&&\exp\left(-\rmi k_0  T\right) \exp\Big[\rmi \sum_j p_j X(Q_j)\Big]\nonumber\\
&&\quad =\exp\Big[\rmi\sum_j p_j X(Q_j-\la k_0)\Big]\exp\left(-\rmi k_0  T\right)\,.\label{bragq}
\ea
When $X(Q_j)=Q_j$, one recovers Eq.~\Eq{braq}. The derivation of the braiding formula \Eq{brak} for $\kappa$-Minkowski spacetime was simple enough, but we can replicate it now in just one line. Noting that the coordinates
\be\label{qmapk} 
Q_j := \ln X_j
\ee
are canonical in the orthant $X_i>0$, one has $X_i=\exp(Q_i)=X(Q_i)$ and Eq.~\Eq{brak} is recovered after analytic continuation to the whole $(D-1)$-dimensional space.

Equation \Eq{bragq} is the composition law of phases for \emph{any} algebra of coordinates $T$ and $X_i$ which can be mapped onto the canonical algebra via an invertible transformation
\be\label{cama}
X_i=X(Q_i)\,,\qquad Q_i=Q_i(X_i)= X^{-1}(X_i)\,.
\ee
For some special cases, phases obey a group composition law. This happens for all Lie-algebraic (linear) noncommutative spaces, where the direct construction works, and for canonically mapped algebras where the mapping \Eq{cama} is linear [$X(Q+a)=X(Q)+X(a)$, Heisenberg algebra], or such that $X(Q+a)=X(Q) g(a)$ for some function $g$. The latter case is $\kappa$-Minkowski spacetime, but not only: for instance, if $Q_i={\rm arcsin}(X_i)$ (up to scales) and space is compact with $\la$-dependent period (discrete momentum space), then the phases in Eq.~\Eq{bragq} remain linear in $X_i$.

Outside these cases, however, the braiding formula tells us that phases do not satisfy a group law, and one should change the Weyl map \Eq{Weyl} or, in other words, the definition of ``plane wave.'' The Weyl mapping may be not unique, but one such prescription exists for all algebras related to the Heisenberg algebra via the canonical transformation \Eq{cama}. The natural plane waves of these algebras are
\be\label{kplaw}
w_k(X):= \exp\left[-\rmi k_0 T+ i \sum_{j=1}^{D-1} k_j Q(X_j)\right]\,,
\ee
the Weyl map is $\Om_q$, 
\be\label{caom2}
\Om_q(f)=\frac{1}{(2\pi)^D}\int_{-\infty}^{+\infty}\rmd^D k\,w_k(X)\tilde f(k)\,,
\ee
and the star product is the canonical $*$-product \Eq{starq}.

At this point, the calculation of the cyclicity-invariant measure associated with a given algebra is immediately obtained via the Jacobian method. For $\kappa$-Minkowski spacetime, the coordinate transformation $x_j=\exp{(q_j)}$ provides an isomorphism between canonical space and $\kappa$-Minkowski space in the first orthant ($x_j>0$) \cite{AAAD}. Then, the Jacobian of this transformation is just the particular solution \Eq{kiacob}, $J(x)=v_\k({\bf x})$.

In the case of $\kappa$-Minkowski spacetime, there exist inequivalent *-products in one-to-one correspondence. In fact, we have seen that a brute-force calculation of the cyclicity-invariant measure \Eq{kiacob} stems from three different Weyl maps, adopting the time-to-the-right, the time-symmetrized, and the symmetric normal ordering [Eqs.~\Eq{R}, \Eq{T}, and \Eq{S}]. Now we see that a fourth alternative is the canonical *-product $*_q$, defined on the plane waves \Eq{kplaw} with the Weyl map \Eq{caom2}. Choosing different Weyl maps corresponds to changing the basis on which functionals are expanded. When we shall consider general nonlinear algebras, the most convenient choice will be by far $*_q$ and the Weyl map \Eq{caom2}. Since we have not been able to prescribe alternative quantizations, we do not know whether this choice is unique.


\section{Noncommutative fractal spacetimes}\label{nfs}

Having prepared all the necessary ingredients, we are ready to present the main physical results of this paper. Since the geometry of multifractal spacetimes changes with the scale, it is natural to expect that any relation between these and noncommutative spaces with a given coordinate algebra will be valid at a fixed scale. At different scales, the mapping will also change.


\subsection{Mapping in the near-boundary regime}\label{bou}

At the smallest scale probed by fractional models, there is a simple correspondence with $\k$-Minkowski spacetime. If the time direction is nonfractional and $\a_0=1$, the measure weight \Eq{noncor} in the boundary-effect regime reads
\be\label{noncor2}
v_{\rm BE}({\bf x}) = \prod_{j=1}^{D-1}x_j^{-1}=v_\k({\bf x})\,,
\ee
which coincides with the $\k$-Minkowski cyclicity-preserving measure. The only difference is in the support: while $v_{\rm BE}$ lives in the first orthant, the weight $v_\k$ was found in the same region but then analytically continued
to the whole space $\mathbb{R}^{D-1}$. In turn, analytic continuation was due to the particular canonical mapping \Eq{qmapk} employed, which was well defined only in the first orthant.

The fundamental scale of $\kappa$-Minkowski (what noncommutativists would call ``the Planck length''\footnote{Note that at this stage this is nothing more than a (reasonable) choice.}) is then identified with $\ell_\infty$:
\be
\ell_\infty=\lp\,.
\ee
Equations \Eq{dh} and \Eq{dstar} imply that the critical point with lowest integer Hausdorff dimension in a $D=4$ ambient space with integer time has $\dh=3$. This may be in apparent agreement with the fact that the spectral dimension of $\kappa$-Minkowski spacetime is 3 \cite{Ben08}, but there are two caveats one should not overlook. First, the result of \cite{Ben08} relies on the noncyclic-invariant measure $v_\kappa=1$. Second, in anisotropic models with $\a\neq\a_0=1$ the spectral and Hausdorff dimensions are different. In order to get a correct comparison, one should compute the spectral dimension with the method of \cite{Ben08} for the cyclicity-preserving measure, and verify that it coincides with $\ds=2$ as given by Eq.~\Eq{ds} ($D=4$ and $\a=1/2$). Since the calculation of the spectral dimension follows the same method of the return probability, the two results should agree; we omit, however, a detailed calculation.

Note that, in order to establish the connection between the two measures, we necessarily had to start with a complex measure with log-oscillations. Indeed, in order to recover the measure \Eq{noncor2} at scales $\ell\gtrsim \ell_*$, i.e., in a real-order fractional action with measure $\vr_\a$ and $\a_0=1$, one should send $\a$ to zero and formally keep the leading term in the expansion of the measure weight (now coefficients are fully reinstated) $v_\a(x^\mu)=(x^\mu)^{\a-1}/\Gamma(\a)\sim \a/x^\mu$, getting
\be\label{fokmin}
\rmd\vr_\a\ \stackrel{\a\to0}{\sim}\ \a^{D-1} v_{\rm BE}(x)\,\rmd^D x\,.
\ee
In the pure multifractional scenario, this expression has little physical meaning. First of all, the correct limit $\a\to 0$ of the measure weight should be taken in the sense of distributions, and it reads $v_\a\sim\delta(x)$ (e.g., \cite{frc1}); in this respect, the formal inverse-power limit \Eq{fokmin} is at least doubtful. One could simply ignore this issue, take Eq.\ \Eq{fokmin} at face value, and absorb the vanishing constant $\a^{D-1}$ into a new normalization for the action, so as to obtain a finite limit when $\a$ goes to zero. However, from the geometric considerations constraining the range of $\a$ (validity of the triangle inequality), models with $\a<1/2$ have a problematic interpretation. This is all the more the case for $\a=0$, which has a pathological geometric structure: the associated measure describes a spacetime with effectively zero spatial dimensions.

On the other hand, all these problems disappear in complex multifractional models. In that case, there is no need to take the $\a\to 0$ limit: The measure weight $v_{\rm BE}$ is well defined at scales $\ell\sim\ell_\infty$, where it is just the asymptotic limit of the log-oscillating measure in the boundary-effect regime. As we mentioned above, in that regime the Hausdorff and spectral dimensions take well-defined, nonzero values determined by the fractional charge frozen at $\a(\ell_*)$.


\subsection{Mapping in the multifractional regime}\label{mufr}

In the real-order multifractional regime $\ell_\om\ll\ell\lesssim \ell_*$, the effective measure weight is the average \Eq{mfrm} of the measure over a log-period, Eq.~\Eq{avev}. If $\a_0=1$,
\be\label{uffa}
v_\a({\bf x}) = \prod_{j=1}^{D-1} x_j^{\a-1}\,.
\ee 
At any given $\a$, we demonstrate that this measure is also obtained, if one imposes the cyclicity condition, in a noncommutative spacetime endowed with the nonlinear algebra
\be\label{alg}
[X_i,T]=\rmi\la X_i^{1-\a}\,,\qquad [X_i,X_j]=0\,,
\ee
where $\la>0$ is real. For $\a=1$ one obtains canonical spacetime and the Heisenberg algebra, Eq.~\Eq{cano}, unless one changes the physical interpretation and imposes commutative space in the infrared, where the dimensional flow ends. In this case, it is sufficient to redefine the constant in the algebra as $\la\to (1-\a)\la$. Anyway, for $\a=0$ one gets $\kappa$-Minkowski spacetime, Eq.~\Eq{km}. The cyclicity-inducing measure(s) will not depend on $\la$.

One can check that if the algebra \Eq{alg} holds, then
\be\label{qX}
Q_i := \frac{X_i^{\a}}{\a}
\ee
are canonical coordinates in the first orthant.\footnote{This is the reason why the geometric coordinates \Eq{fpotra} were denoted as $q^\mu$ in \cite{fra4,frc1,frc2}.} In fact, $[Q_i,Q_j]=0$ from the second relation in \Eq{alg}. To prove the remaining commutator, we need the commutator of $T$ with an arbitrary, possibly noninteger power $\b$ of $X_i$. For an integer power $\b=N$, this commutator would read (index $i$ omitted)
\be\nonumber
[X^N,T]=\sum_{n=0}^{N-1} X^{N-1-n}[X,T] X^n\,,
\ee
from which a generalization to the continuum follows:
\be
[X^\b,T]=\int_{0}^{\b}\rmd s\, X^{\b-1-s}[X,T] X^s = \rmi\b\la X^{\b-\a}\,,
\ee
where the last equality holds for the algebra \Eq{alg}. Then, setting $\b=\a$ we get
\be
[Q_i,T]=\frac{1}{\a}[X_i^\a,T]=\rmi\la\,.
\ee
Notice that in the limit $\a\to 0$, and after removing an immaterial divergent constant, Eq.~\Eq{qX} gives the known mapping from canonical to $\kappa$-Minkowski coordinates, Eq.~\Eq{qmapk}. This is the reason why we put a factor $1/\a$ in \Eq{qX}.

From the discussion in Sec.~\ref{braid} and Eq.~\Eq{bragq}, one sees that phases are not natural plane waves in fractional noncommutative spacetime described by the algebra \Eq{alg}:
\ba
&&\exp(-\rmi k_0  T)\,\exp(\rmi p_j X^j)\nonumber\\
&&\quad =\exp\Big[\rmi\sum_j p_j (\a X^\a_j-\a\la k_0)^{1/\a}\Big]\,\exp(-\rmi k_0 T).\label{branl}
\ea
In the limits $\a\to1$ and $\a\to 0$, one correctly recovers Eqs.~\Eq{braq} and \Eq{brak}, respectively. However,
one can construct a quantum field theory on these spacetimes by employing the canonical Weyl map and $*$-product. 
The effective spacetime measure is then found immediately via the Jacobian trick. For general $\a$, the Jacobian associated with the coordinate transformation \Eq{qX} is
\be
J(x) = \left|\frac{\p q}{\p x}\right| =\prod_{i=1}^{D-1} |x_i|^{\a-1}\,,
\ee
and by the known calculation in canonical space, this is also a particular solution for a cyclicity-preserving measure. It coincides with Eq.~\Eq{uffa} and constitutes the final result. 

We have thus shown that we can map the measures appearing in field theories on multifractional spacetimes $S(\phi)=\int \rmd \vr(x)\,\cL(\phi)\, = \int d^Dx\, v_\alpha(x)\,\cL[\phi(x)]$ and in field theories on spacetimes characterized by a nonlinear noncommutative algebra of coordinates (\ref{alg}) $\mathcal{I}(\phi) = \int d^Dx\, v_\alpha(x)\, F[\phi(x)]$. We have not, however, shown that this extends to a map of {\it field theories} that include also the algebra of fields and the corresponding dynamical action principles. This was beyond our present goal.

As already mentioned, spacetimes associated with this measure have Hausdorff and spectral dimension equal to, respectively, $\dh=1+(D-1)\a$ and $\ds=D\a$ \cite{frc1}. Since $\dh\neq\ds$, diffusion processes taking place therein are anomalous.

One may be concerned that the series \Eq{eg} does not exist for $X_i$ defined as in Eq.~\Eq{qX}. The canonical mapping is valid even in the case where \Eq{eg} is formal, but for the sake of completeness we note that we can take another mapping which gives exactly the same results:
\ba
Q_i &:=& \frac{X_i^{\a}}{\a}-1\,,\\
X_i &=& (1+\a Q_i)^{1/\a}=\sum_{n=0}^{+\infty}\binom{\a^{-1}}{n}(\a Q_i)^n\,,\label{qX2}
\ea
where
\be
\binom{\a^{-1}}{n}=\frac{\Gamma(1+\a^{-1})}{n!\Gamma(\a^{-1}-n+1)}\,.
\ee
Obviously, these new coordinates are canonical and the series is well defined since $0<\a\leq 1$.


\subsection{Representations}

In order to find the Hilbert space whereon the fractional algebra \Eq{alg} acts, we must find an explicit representation. Here we are interested in mathematical realizations of a representation on an abstract functional space, not in any specific physical realization.

Because of the fact that the fractional Leibniz rule is considerably more complicated than the integer case, we were unable to find representations based upon fractional derivatives. Nonetheless, a representation associated with the canonical mapping may be inferred from the Heisenberg algebra. Letting
\be\label{reprq}
T:= -\rmi\la\sum_{j=1}^{D-1}\frac{\rmd}{\rmd s_j}\,,\qquad Q_i:= s_i\,,
\ee
one has 
\be\nonumber
[Q_j,T] f(s)=\rmi\la\left(\frac{\rmd}{\rmd s_j}s_j\right)f(s)= \rmi\la f(s)\,.
\ee
Then, in fractional Minkowski spacetime the canonical representation is
\be\label{repr0}
T:= -\rmi\la\sum_{j=1}^{D-1}\frac{\rmd}{\rmd s_j}\,,\qquad X_i:= (\a s_i)^{1/\a}\,.
\ee
In fact,
\ba
[X_j,T]f(s) &=&  \rmi\la\left[\frac{\rmd}{\rmd s_j}(\a s_j)^{1/\a}\right]f(s)\nonumber\\
        &=& \rmi\la(\a s_j)^{1/\a-1} f(s)\nonumber\\
        &=&  \rmi\la X_j^{1-\a}f(s)\nonumber\,.
\ea
Notice that the operator $T$ is a plain derivative, i.e., the generator of translations. By definition, ``plane waves'' are the eigenfunctions of the Laplacian, and Eq.~\Eq{repr0} tells us that the plane waves of the quantum theory are phases in $Q_i(X)$. This is in agreement with the canonical mapping picture.

\section{Conclusions}\label{conc}

We have provided a connection between multifractional and noncommutative spacetimes based on the properties of their nontrivial integration measures. Near the fundamental scale $\ell_\infty$, we found a correspondence between the fractional measure and the cyclicity-inducing measure of $\kappa$-Minkowski spacetime with the identification $\ell_\infty\sim\lp\sim1/\kappa$. For scales in the multifractional regime $\ell_\infty<\ell\lesssim\ell_*$ we showed that the fractional measure is connected to the cyclicity-inducing measure of a family of fractional noncommutative spacetimes whose fractional index $\alpha$ can be seen as a parameter which interpolates between canonical ($\alpha=1$) and $\kappa$-Minkowski ($\alpha=0$) noncommutative spacetime. 

A couple of technical points may throw some light on the physical picture. First, while fractional measures have support in the first orthant, noncommutative spaces typically extend to the whole embedding. The canonical mapping for $\k$-Minkowski coordinates was determined only in the first orthant, too, but then one can analytically continue \cite{AAAD} and get the correct measure, with absolute values. The fact that we have an analytic obstruction in the fractional case might be significant for clarifying the fractal-noncommutative duality. 

Second, cyclic-invariant measures are not unique, and the duality was determined between fractional measures on one hand and the particular solution $v=J$ on the other hand, corresponding to the Lebesgue canonical measure $v_c=1$. We do not know whether other nontrivial solutions exist with a fractal interpretation, but a negative answer would imply that the canonical mapping is a far stricter guiding principle than expected. The lack of alternative constructions, not passing through the canonical mapping, leaves this matter undecided.

Two other pressing issues remain open. As we discussed, multifractal spacetimes exhibit different types of symmetries at different scales, ranging from discrete scale invariance to nonlinear realizations of Poincar\'e transformations. Also on the noncommutative side, in the simpler cases of $\kappa$-Minkowski and canonical noncommutativity, there exist notions of {\it deformed} Poincar\'e transformations. For $\kappa$-Minkowski, they can also be understood as nonlinear deformations of the ordinary ones. A natural question is whether we can establish a connection between the various symmetry structures on both sides. 

For the specific case of $\kappa$-Minkowski spacetime, another interesting problem is to calculate the Hausdorff and spectral dimensions in the presence of a cyclicity-inducing measure. A measure theory can be defined in order to generalize the Hausdorff and spectral dimensions to noncommutative spaces \cite{GI1,GI2,GI3,Man06}. However, the standard techniques employed in \cite{Ben08} (for the $\k$-Minkowski Laplacian and the trivial Lebesgue measure) and \cite{frc1,frc2} (for classical fractional spacetimes with fractional or integer Laplacians) should be sufficient to determine $\dh$ and $\ds$. It would also be interesting to reproduce the study \cite{Ben08} of the spectral dimension in $\kappa$-Minkowski in the case of the more general nonlinear noncommutative algebra (\ref{alg}). This will be the subject of future work.

Finally, one would like to extend the correspondence between fractional and noncommutative spacetimes that has been investigated in this paper to the level of the field theories defined thereon, mapping both the kinematics of fields defined on such spaces (both classical and quantum) and their dynamics. 

Concerning this, as we mentioned, the respective role of cyclicity and breaking of symmetries in noncommutative field theories should be further elucidated. From the perspective of noncommutative geometry, the effective measure of $\kappa$-Minkowski spacetime is particularly problematic. On one hand, the classical (commutative) limit $\lambda\sim\lp\to 0$ is sensible in the intermediate steps leading to the cyclicity equations, Eqs.\ \Eq{cycleq1}, but $\lambda$ completely disappears in the equations themselves. Then, the measure \Eq{kiacob} bears no dependence on the quantum length scale, and it is not obvious how to recover the commutative limit in classical-coordinate space. On the other hand, the class of cyclicity-preserving measures does not enjoy all (or any of) the $\kappa$-Poincar\'e symmetries, which should play a role at least as important as cyclicity in the construction of a field theory. None of these problems has ever been solved by purely noncommutative tools. We can interpret this status of affairs in two radically different ways. If symmetry requirements were to be taken more seriously, then one should abandon cyclicity and, as an indirect consequence, revise or abandon the mapping with fractional theories. However, an interesting and more optimistic change of perspective, worth exploring in the future, is to promote the fractional side of the mapping as fundamental. In this case, the noncommutative side is only an effective description of a theory where geometry and symmetries are well defined at all scales, including the infrared limit. Then, the measure problem in $\kappa$-Minkowski is solved by recognizing $\kappa$-Minkowski as an asymptotic, not exact, geometry endowed with accidental quantum symmetries. From the fractional side, one begins with an oscillating measure $\vr$ featuring the fundamental length $\lp$. At large scales, $\vr$ becomes the ordinary Lebesgue measure; this is \emph{not} achieved by the formal limit $\lp\to 0$, but by a nontrivial dimensional flow typical of multifractal geometry. At scales of order of the Planck length, the measure tends to the cyclicity-preserving measure of the noncommutative calculation, and the dependence on $\lp$ is factorized in a constant prefactor that is simply ignored in the latter case.

The preservation of symmetries in theories where these are given by quantum groups (e.g., $\kappa$-Minkowski and canonical noncommutativity) seems to require the introduction of a nontrivial {\it braiding} for field operators and a nontrivial statistics \cite{oeckl,sasaisasakura,balachandran,Arzano:2007ef,Arzano:2008bt}. Despite its obvious importance, this issue is not fully understood in the noncommutative field theory setting (in particular, $\kappa$-Minkowski spacetime) and it has been noticed only recently in fundamental quantum gravity models like group field theory \cite{GFTdiffeos}. It should certainly be investigated further both for the noncommutative algebra (\ref{alg}) introduced in this paper and in the fractional spacetime context.


\begin{acknowledgments}
M.A.\ is supported by the EU Marie Curie Actions under a Marie Curie Intra-European Fellowship. G.C., D.O., and M.S.\ thank C.~Chryssomalakos, F.~Lizzi, and R.~Loll for useful discussions. D.O.\ gratefully acknowledges support from the A.~von Humboldt Stiftung through a Sofja Kovalevskaja Award. M.S.\ acknowledges support from the University of Catania, Italy, within the International Internship Programme, and a grant from INFN.
\end{acknowledgments}


\begin{thebibliography}{99}

\bibitem{Ori06} D.~Oriti, 
 in {\it Approaches to Quantum Gravity}, edited by D.~Oriti (Cambridge University Press, Cambridge, England, 2009) [\oarX{gr-qc/0607032}].
\bibitem{Ori11} D.~Oriti, in {\it Foundations of space and time}, edited by G.\ Ellis, J.\ Murugan, and A.\ Weltman (Cambridge University Press, Cambridge, England, 2011)
\bibitem{Rov06} C.~Rovelli, {\it Quantum Gravity} (Cambridge University Press, Cambridge, England, 2006).
\bibitem{Thi07} T.~Thiemann, {\it Modern canonical quantum General Relativity} (Cambridge University Press, Cambridge, England, 2007).
\bibitem{Ori01} D.~Oriti, 
 \doin{10.1088/0034-4885/64/12/203}{Rep.\ Prog.\ Phys.\ \textbf{64}, 1703 (2001)} [\oarX{gr-qc/0106091}].
 \bibitem{Per03} A.~Perez, 
 \doin{10.1088/0264-9381/20/6/202}{Classical Quantum Gravity \textbf{20}, R43 (2003)} [\oarX{gr-qc/0301113}].
\bibitem{Wil2}  H.~Hamber, \doin{10.1007/s10714-009-0769-y}{Gen.\ Relativ.\ Gravit.\ {\bf 41}, 817 (2009)} [\arX{0901.0964}].
\bibitem{Wil3}  R.M.~Williams, 
\doin{10.1016/S0920-5632(97)00355-1}{Nucl.\ Phys.\ B, Proc.\ Suppl.\ {\bf 57}, 73 (1997)} [\oarX{gr-qc/9702006}].
\bibitem{lol08} R.~Loll, 
 \doin{10.1088/0264-9381/25/11/114006}{Classical Quantum Gravity {\bf 25}, 114006 (2008)} [\arX{0711.0273}].
\bibitem{lol2}  J. Ambj{\o}rn, A.~Gorlich, J.~Jurkiewicz, and R.~Loll, \arX{1007.2560}.
\bibitem{lol3}  J. Ambj{\o}rn, J. Jurkiewicz, and R.~Loll, \doin{10.1007/978-3-642-11897-5_2}{Lect.\ Notes Phys.\ {\bf 807}, 59 (2010)} [\arX{0906.3947}].
\bibitem{Nie06} M.~Niedermaier, 
\doin{10.1088/0264-9381/24/18/R01}{Classical Quantum Gravity {\bf 24}, R171 (2007)} [\oarX{gr-qc/0610018}].
\bibitem{NiR}   M.~Niedermaier and M.~Reuter, 
 \href{http://www.livingreviews.org/lrr-2006-5}{Living Rev.\ Relativity {\bf 9}, 5 (2006)}. 
\bibitem{ReS4}  M.~Reuter and F.~Saueressig, 
\arX{0708.1317}.
\bibitem{CPR}   A.~Codello, R.~Percacci, and C.~Rahmede, 
\doin{10.1016/j.aop.2008.08.008}{Ann.\ Phys.\ (N.Y.) {\bf 324}, 414 (2009)} [\arX{0805.2909}].
\bibitem{Sot10} T.P.~Sotiriou,  
  \doin{10.1088/1742-6596/283/1/012034}{J.\ Phys.\ Conf.\ Ser.\ {\bf 283}, 012034 (2011)} [\arX{1010.3218}].
\bibitem{Muk10} S.~Mukohyama,  
  \doin{10.1088/0264-9381/27/22/223101}{Classical Quantum Gravity {\bf 27}, 223101 (2010)} [\arX{1007.5199}].

\bibitem{Madore} J.~Madore, {\it An introduction to noncommutative geometry and its physical applications} (Cambridge University Press, Cambridge, England, 1999).
\bibitem{Szabo} R.J.~Szabo, 
 \doin{10.1016/S0370-1573(03)00059-0}{Phys.\ Rep.\ {\bf 378}, 207 (2003)} [\oarX{hep-th/0109162}].
\bibitem{Connes} A.~Connes, {\it Noncommutative geometry} (Academic Press, San Diego, 2004).
\bibitem{ADKLW} P.~Aschieri, M.~Dimitrijevic, P.~Kulish, F.~Lizzi, and J.~Wess, {\it Noncommutative spacetimes} (Springer-Verlag, Berlin, 2009).
\bibitem{SW} N.~Seiberg and E.~Witten, 
 \doin{10.1088/1126-6708/1999/09/032}{J.\ High Energy Phys.\ 09 (1999) 032} [\oarX{hep-th/9908142}].
\bibitem{MST} A.~Matusis, L.~Susskind, and N.~Toumbas, \doin{10.1088/1126-6708/2000/12/002}{J.\ High Energy Phys.\ 12 (2000) 002} [\oarX{hep-th/0002075}].
\bibitem{ACSS} G.~Amelino-Camelia, L.~Smolin, and A.~Starodubtsev, \doin{10.1088/0264-9381/21/13/002}{Classical Quantum Gravity {\bf 21}, 3095 (2004)} [\oarX{hep-th/0306134}].
\bibitem{FKGS} L.~Freidel, J.~Kowalski-Glikman, and L.~Smolin, \doin{10.1103/PhysRevD.69.044001}{Phys.\ Rev.\ D {\bf 69}, 044001 (2004)} [\oarX{hep-th/0307085}].

\bibitem{Doplicher:1994zv}
  S.~Doplicher, K.~Fredenhagen, and J.E.~Roberts,
  \doin{10.1016/0370-2693(94)90940-7}{Phys.\ Lett.\ B {\bf 331}, 39 (1994)}.
\bibitem{Doplicher:1994tu}
  S.~Doplicher, K.~Fredenhagen, and J.E.~Roberts,
  \doin{10.1007/BF02104515}{Commun.\ Math.\ Phys.\ {\bf 172}, 187 (1995)}.
  [\oarX{hep-th/0303037}].
\bibitem{Majid:1994cy}
  S.~Majid and H.~Ruegg,
  \doin{10.1016/0370-2693(94)90699-8}{Phys.\ Lett.\ B {\bf 334}, 348 (1994)} [\oarX{hep-th/9405107}].
\bibitem{Agostini:2003vg}
  A.~Agostini, G.~Amelino-Camelia, and F.~D'Andrea,
  \doin{10.1142/S0217751X04020919}{Int.\ J.\ Mod.\ Phys.\ A {\bf 19}, 5187 (2004)} [\oarX{hep-th/0306013}].
\bibitem{Matschull:1997du}
  H.-J.~Matschull and M.~Welling,
  \doin{10.1088/0264-9381/15/10/008}{Classical Quantum Gravity {\bf 15}, 2981 (1998)} [\oarX{gr-qc/9708054}].
\bibitem{Deser:1983tn}
  S.~Deser, R.~Jackiw, and G.~'t Hooft,
  \doin{10.1016/0003-4916(84)90085-X }{Ann.\ Phys.\ (N.Y.) {\bf 152}, 220 (1984)}.
\bibitem{ArzanoAlesci}
M.~Arzano and E.~Alesci, \arX{1108.1507}.

\bibitem{fra1}  G.~Calcagni, 
\doin{10.1103/PhysRevLett.104.251301}{Phys.\ Rev.\ Lett.\ {\bf 104}, 251301 (2010)} [\arX{0912.3142}].
\bibitem{fra2}  G.~Calcagni, 
 \doin{10.1007/JHEP03(2010)120}{J.\ High Energy Phys.\ 03 (2010) 120} [\arX{1001.0571}].
\bibitem{fra3}  G.~Calcagni, 
\doin{10.1016/j.physletb.2011.01.063}{Phys.\ Lett.\ B {\bf 697}, 251 (2011)} [\arX{1012.1244}].
\bibitem{fra4}  G.~Calcagni, 
 \doin{10.1103/PhysRevD.84.061501}{Phys.\ Rev.\ D {\bf 84}, 061501(R) (2011)} [\arX{1106.0295}].
\bibitem{frc1}  G.~Calcagni, \arX{1106.5787}.
\bibitem{frc2}  G.~Calcagni, \arX{1107.5041}.

%
\bibitem{AAAD} A.~Agostini, G.~Amelino-Camelia, M.~Arzano and F.~D'Andrea, \doin{10.1142/S0217751X06031077}{Int.\ J.\ Mod.\ Phys.\ A {\bf 21}, 3133 (2006)} [\oarX{hep-th/0407227}].
\bibitem{Ben08} D.~Benedetti, 
 \doin{10.1103/PhysRevLett.102.111303}{Phys.\ Rev.\ Lett.\ {\bf 102}, 111303 (2009)} [\arX{0811.1396}].

\bibitem{Fal03} K.~Falconer, \textit{Fractal geometry} (Wiley, New York, 2003).
\bibitem{Har01} D.~Harte, {\it Multifractals: theory and applications} (Chapman \& Hall/CRC, Boca Raton, 2001).
\bibitem{Pod99} I.~Podlubny, \emph{Fractional differential equations} (Academic Press, San Diego, 1999).
\bibitem{KST}   A.A.~Kilbas, H.M.~Srivastava, and J.J.~Trujillo, \emph{Theory and applications of fractional differential equations} (Elsevier, Amsterdam, 2006).
\bibitem{RLWQ}  F.-Y.~Ren, J.-R.~Liang, X.-T.~Wang, and W.-Y.~Qiu, 
 \doin{10.1016/S0960-0779(02)00211-4}{Chaos Solitons Fractals {\bf 16}, 107 (2003)}.
\bibitem{NLM}   R.R.~Nigmatullin and A.~Le M\'ehaut\'e, 
\doin{10.1016/j.jnoncrysol.2005.05.035}{J.\ Non-Cryst.\ Solids {\bf 351}, 2888 (2005)}.
\bibitem{Sor98} D.~Sornette, \doin{10.1016/S0370-1573(97)00076-8}{Phys.\ Rep.\ {\bf 297}, 239 (1998)} [\oarX{cond-mat/9707012}].

%
\bibitem{Weyl} H.~Weyl, 
\doin{10.1007/BF02055756}{Z.\ Phys.\ {\bf 46}, 1 (1927)}; {\it The theory of groups and quantum mechanincs} (Dover, New York, 1931).
\bibitem{Gutt} S.~Gutt, \oarX{math/0003107}.

\bibitem{MSSW}  J.~Madore, S.~Schraml, P.~Schupp, and J.~Wess,  
  \doin{10.1007/s100520050012}{Eur.\ Phys.\ J.\ C {\bf 16}, 161 (2000)} [\oarX{hep-th/0001203}].

\bibitem{Kosinski:1999dw}
  P.~Kosi\'nski, J.~Lukierski, and P.~Ma\'slanka,
  \doin{10.1023/A:1022821310096}{Czech.\ J.\ Phys.\ {\bf 50}, 1283 (2000)} [\oarX{hep-th/0009120}].
  
\bibitem{Agostini:2002de}
  A.~Agostini, F.~Lizzi, and A.~Zampini,
  \doin{10.1142/S021773230200871X}{Mod.\ Phys.\ Lett.\ A {\bf 17}, 2105 (2002)} [\oarX{hep-th/0209174}].

\bibitem{Mil72} W.~Miller, {\it Symmetry groups and their applications} (Academic Press, New York, 1972).

\bibitem{Chaichian:2004yh}
  M.~Chaichian, P.~Pre\v{s}najder, and A.~Tureanu,
  \doin{10.1103/PhysRevLett.94.151602}{Phys.\ Rev.\ Lett.\ {\bf 94}, 151602 (2005)} [\oarX{hep-th/0409096}].

\bibitem{Balachandran:2007vx}
  A.P.~Balachandran, A.~Pinzul, and B.A.~Qureshi,
  \doin{10.1103/PhysRevD.77.025021}{Phys.\ Rev.\ D {\bf 77}, 025021 (2008)} [\arX{0708.1779}].

\bibitem{DJMTWW} M.~Dimitrijevic, L.~Jonke, L.~M\"oller, E.~Tsouchnika, J.~Wess. and M.~Wohlgenannt, \doin{10.1140/epjc/s2003-01309-y}{Eur.\ Phys.\ J.\ C {\bf 31}, 129 (2003)} [\oarX{hep-th/0307149}].
\bibitem{eterawinston} W.~Fairbairn and E.~Livine, \doin{10.1088/0264-9381/24/20/021}{Classical Quantum Gravity \textbf{24}, 5277 (2007)} [\oarX{gr-qc/0702125}].
\bibitem{noi} F.~Girelli, E.~Livine, and D.~Oriti, \doin{10.1103/PhysRevD.81.024015}{Phys.\ Rev.\ D {\bf 81}, 024015 (2010)} [\arX{0903.3475}].
\bibitem{danieleEmerg} D.~Oriti, \doin{10.1088/1742-6596/174/1/012047}{J.\ Phys.\ Conf.\ Ser.\ {\bf 174}, 012047 (2009)} [\arX{0903.3970}].
\bibitem{LRZ} J.~Lukierski, H.~Ruegg, and W.J.~Zakrzewski, \doin{10.1006/aphy.1995.1092}{Annals Phys.\ {\bf 243}, 90 (1995)} [\oarX{hep-th/9312153}].
\bibitem{LNRT} J.~Lukierski, A.~Nowicki, H.~Ruegg, and V.N.~Tolstoy, \doin{10.1016/0370-2693(91)90358-W}{Phys.\ Lett.\ B {\bf 264}, 331 (1991)}.

\bibitem{AmelinoCamelia:2000mn}
  G.~Amelino-Camelia,
  \doin{10.1142/S0218271802001330}{Int.\ J.\ Mod.\ Phys.\ D {\bf 11}, 35 (2002)} [\oarX{gr-qc/0012051}].

\bibitem{KowalskiGlikman:2002we}
  J.~Kowalski-Glikman and S.~Nowak,
  \doin{10.1016/S0370-2693(02)02063-4}{Phys.\ Lett.\ B {\bf 539}, 126 (2002)} [\oarX{hep-th/0203040}].

\bibitem{CamArz} G.~Amelino-Camelia and M.~Arzano, \doin{10.1103/PhysRevD.65.084044}{Phys.\ Rev.\ D {\bf 65}, 084044  (2002)} [\oarX{hep-th/0105120}].
\bibitem{Arzano:2007nx}
  M.~Arzano,
  \doin{10.1103/PhysRevD.77.025013}{Phys.\ Rev.\ D {\bf 77}, 025013 (2008)} [\arX{0710.1083}].
  
\bibitem{AmelinoCamelia:2011bm}
  G.~Amelino-Camelia, L.~Freidel, J.~Kowalski-Glikman, and L.~Smolin,
  \doin{10.1103/PhysRevD.84.084010}{Phys.\ Rev.\ D {\bf 84}, 084010 (2011)} [\arX{1101.0931}].
 
 \bibitem{AmelinoArzRelLoc}
     G.~Amelino-Camelia, M.~Arzano, J.~Kowalski-Glikman, G.~Rosati, G.~Trevisan,
  [\arX{1107.1724}].
   
\bibitem{Arzano:2010jw}
  M.~Arzano,
  \doin{10.1103/PhysRevD.83.025025}{Phys.\ Rev.\ D {\bf 83}, 025025 (2011)} [\arX{1009.1097}].

\bibitem{Arzano:2007ef}
  M.~Arzano and A.~Marcian\`o,
  \doin{10.1103/PhysRevD.76.125005}{Phys.\ Rev.\ D {\bf 76}, 125005 (2007)} [\arX{0707.1329}].
  
\bibitem{Freidel:2007yu}
  L.~Freidel and J.~Kowalski-Glikman,
  \arX{0710.2886}.

\bibitem{Agostini:2006nc}
  A.~Agostini, G.~Amelino-Camelia, M.~Arzano, A.~Marcian\`o, and R.A.~Tacchi,
  \doin{10.1142/S0217732307024024}{Mod.\ Phys.\ Lett.\ A {\bf 22}, 1779 (2007)} [\oarX{hep-th/0607221}].

\bibitem{Arzano:2009ci}
  M.~Arzano, J.~Kowalski-Glikman, and A.~Walkus,
  \doin{10.1088/0264-9381/27/2/025012}{Classical Quantum Gravity {\bf 27}, 025012 (2010)} [\arX{0908.1974}].

\bibitem{DoNe}  M.R.~Douglas and N.A.~Nekrasov, \doin{10.1103/RevModPhys.73.977}{Rev.\ Mod.\ Phys.\ {\bf 73}, 977 (2001)} [\oarX{hep-th/0106048}].
\bibitem{GI1}   D.~Guido and T.~Isola, 
in {\it Mathematical Physics in Mathematics and Physics}, edited by R.~Longo (AMS, Providence, 2001) [\oarX{math/0102209}].
\bibitem{GI2}   D.~Guido and T.~Isola, 
\doin{10.1016/S0022-1236(03)00230-1}{J.\ Funct.\ Anal.\ {\bf 203}, 362 (2003)} [\oarX{math/0202108}].
\bibitem{GI3}   D.~Guido and T.~Isola, 
in {\it Advances in Operator Algebras and Mathematical Physics}, edited by F.~Boca, O.~Bratteli, R.~Longo, and H.~Siedentop (Dundatia Theta, Bucharest, 2006) [\oarX{math/0404295}].
\bibitem{Man06} Y.I.~Manin, \doin{10.1090/S0273-0979-06-01081-0}{Bull.\ Am.\ Math.\ Soc.\ {\bf 43}, 139 (2006)} [\oarX{math/0502016}].
\bibitem{oeckl} R. Oeckl, \doin{10.1016/S0393-0440(01)00016-X}{J.\ Geom.\ Phys.\ {\bf 39}, 233 (2001)} [\oarX{hep-th/0008072}].
\bibitem{sasaisasakura} Y. Sasa and N.~Sasakura, \doin{10.1143/PTP.118.785}{Prog.\ Theor.\ Phys.\ {\bf 118}, 785 (2007)} [\arX{0704.0822}].
\bibitem{balachandran} A.P.~Balachandran, T.R.~Govindarajan, G.~Mangano, A.~Pinzul, B.A.~Qureshi, and S.~Vaidya, \doin{10.1103/PhysRevD.75.045009}{Phys.\ Rev.\ D {\bf 75}, 045009 (2007)} [\oarX{hep-th/0608179}].

\bibitem{Arzano:2008bt}
  M.~Arzano and D.~Benedetti,
  \doin{10.1142/S0217751X09045881}{Int.\ J.\ Mod.\ Phys.\ A {\bf 24}, 4623 (2009)} [\arX{0809.0889}].

\bibitem{GFTdiffeos} A.\ Baratin, F.\ Girelli, and D.\ Oriti, \doin{10.1103/PhysRevD.83.104051}{Phys.\ Rev.\ D {\bf 83}, 104051 (2011)} [\arX{1101.0590}].
\end{thebibliography}
\end{document}